# Miniature insect flight

Mao Sun*

Institute of Fluid Mechanics, Beihang University, Beijing 100191, China

(*m.sun@buaa.edu.cn)

Approximately half of the existing winged-insect species are of very small size (wing length≈0.3-4 mm); they are referred to as miniature insects. Yet until recently, much of what we know about the mechanics of insect flight was derived from studies on relatively large insects, such as hoverflies, honey bees and hawkmoths. Because of their very small size, many miniature insects fly at a Reynolds number ($Re$) on the order of 10 or less. At such a low $Re$, the viscous effect of the air is very large: A miniature insect moves through the air as would a bumble bee move through mineral oil. Miniature insects must use new flapping mode and new aerodynamic mechanisms to fly. Over the past decade, much work has been done in the study of the mechanics of flight in miniature insects: novel flapping modes have been discovered and new mechanisms of aerodynamic force generation have been revealed; progress has also been made on the fluid-mechanics related flight problems, such as flight power requirements and flight dynamic stability. This article reviews these developments and discusses potential future directions.

## I. INTRODUCTION

The size of winged insects varies greatly (Dudley, 2000; Polilov, 2015). The insects we often see around us, e.g. hoverflies, honey bees, hawkmoths, are relatively large ones; their wing length ($R$) ranges from approximately 5-50 mm (Dudley, 2000). But many winged insects are of very small size, $R$ being approximately 0.3-4 mm (Dudley, 2000; Polilov, 2015) and we rarely notice them. Here we call these very small insects as miniature insects and the relatively large ones as medium and large insects. Fig. 1 gives a comparison between a typical miniature insect (a tiny wasp) and a medium size insect (a dronefly, which has the same size as a honey bee). Close to a half of the existing winged-insect species are of miniature size (Dudley, 2000). But until recently, much of what we know about the mechanics of flight of insects is derived from studies on medium and large insects (Sane, 2016). Before we discuss the flight of miniature insects, let us first have a brief overview of the flight of the medium and large insects (flapping kinematics and aerodynamic mechanisms); more detail reviews can be found in Sane (2003), Wang (2005), Shyy *et al.* (2010) and Shelley and Zhang (2011).

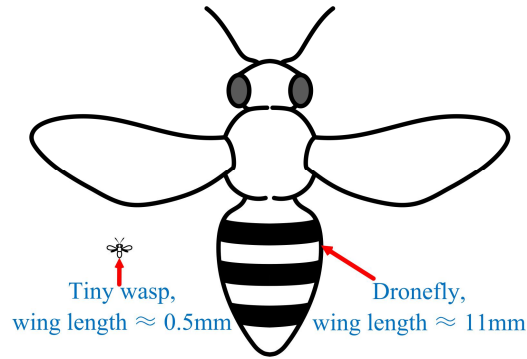

FIG. 1. A diagram showing a comparison between a typical miniature insect (a tiny wasp *Encarsia formosa*) and a medium size insect (a dronefly *Eristalis tenax*, which has the same size as a honey bee).

Medium and large insects in hovering flight usually beat their wings back and forth in an approximately horizontal plane [Fig. 2(a)], referred to as upstroke and downstroke respectively, and the plane in which the wings beat is called stroke plane (Weis-Fogh, 1973; Ellington, 1984*c*; Fry, Sayaman and Dickinson, 2005; Liu and Sun, 2008; Walker, Thomas and Taylor, 2010). In the beginning of an upstroke or downstroke [Fig. 2(b)], the wing accelerates and at the same time pitches down; in the mid-portion of the stroke, the wing moves at approximately constant speed [Fig. 2(b)]; and near the end of the stroke, the wing decelerates and at the same time pitches up [Fig. 2(b)]. In forward flight, the stroke plane tilts forward (Dudley and Ellington, 1900a; Meng and Sun, 2016; Hedrick, Martínez-Blat and Goodman, 2017).

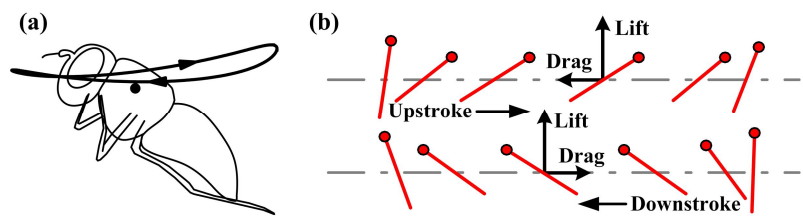

FIG. 2. (a) Stroke diagram shows the wing-tip trajectory (projected onto the symmetrical plane of the insect; the black curve) of a dronefly *Eristalis tenax*, a relatively large insec; a black dot defines the wing-root location on the insect body. (b) The motion of a section of the wing and definition of aerodynamic lift and drag, with dots marking the leading edge.

Although the wing of an insect beats at high frequency (usually above 150 Hz), the velocity of the wing relative to the undisturbed air is small, owing to the small wing length. As a result, the lift coefficient of the wing required to balance the weight is relatively high (Ellington, 1984*a*,*b*,*c*,*d*; Dickinson, Lehman and Sane, 1999); the mean lift coefficient required is around 2 (Liu and Kawachi, 1998; Sun and Du, 2003), about three times as large as that of a cruising airplane. This high lift coefficient cannot be explained by conventional steady-state aerodynamics (which applies to the high Reynolds number and attached flows) and unsteady aerodynamic mechanisms must be operating (Ellington, 1984*c*). Several unsteady mechanisms have been proposed to explain the high aerodynamic-force coefficients. Among them are the 'delayed-stall' mechanism (Ellington *et al.*, 1996; Liu and Kawachi, 1998; Dickinson, Lehman and Sane, 1999; Bomphrey *et al.*, 2002; Zhang, 2017), the 'pitching-up rotation' mechanism (Dickinson *et al.*, 1999; Sun and Tang, 2002a; Bomphrey *et al.*, 2017) and the 'fast acceleration' mechanism (Sun and Tang, 2002a; Sane, 2003). The delayed stall mechanism produces high lift by the leading edge vortex (LEV) which attaches to and moves with the wing during an entire up- or downstroke. The 'fast acceleration' mechanism is that a wing in fast acceleration at large angle of attack can produce a large aerodynamic force. The 'pitching rotation' mechanism is that when a wing moves forward and at the same time fast pitch up, large aerodynamic force can be produced.

For many insects in hovering flight, it has been shown that the lift and drag of the wing generally have three peaks in a upstroke or downstroke (Wang, Birch and Dickinson, 2004; Aono, Liang and Liu, 2008; Liu and Sun 2008; Han, Chang and Han, 2016), an example being shown in Fig. 3 (in the figure $C_L$ and $C_D$ are the lift and drag coefficients, respectively; *t* is the time and *T* is the flapping period; $t/T$=0–0.5: upstroke and $t/T$=0.5–1: downstroke). The force peak in the beginning of the upstroke ($t/T$≈0–0.1) or downstroke ($t/T$≈0.5–0.6) is produced by the fast acceleration mechanism. The 'wider' force peak in the mid-portion of the upstroke ($t/T$≈0.1–0.4) or downstroke ($t/T$≈0.6–0.9) is due to the delayed-stall mechanism, i.e. duo to the leading-edge vortex (LEV) attached to and moving with the wing [Fig. 3(c)]. The force peak near the end of the upstroke ($t/T$≈0.4–0.5) or downstroke ($t/T$≈0.9–1) is generated by the pitching-up rotation mechanism. Note that the $C_L$-peak at beginning of a upstroke or downstroke is much smaller than that in the mid-position of the stroke; this is because although the wing is in fast acceleration, it is at the same time performing pitching-down rotation [see Fig. (2b)]. The $C_L$-peak near the end of the upstroke and downstroke is also smaller; the reason for this is that although the wing is in pitching-up rotation, it is also in fast deceleration.

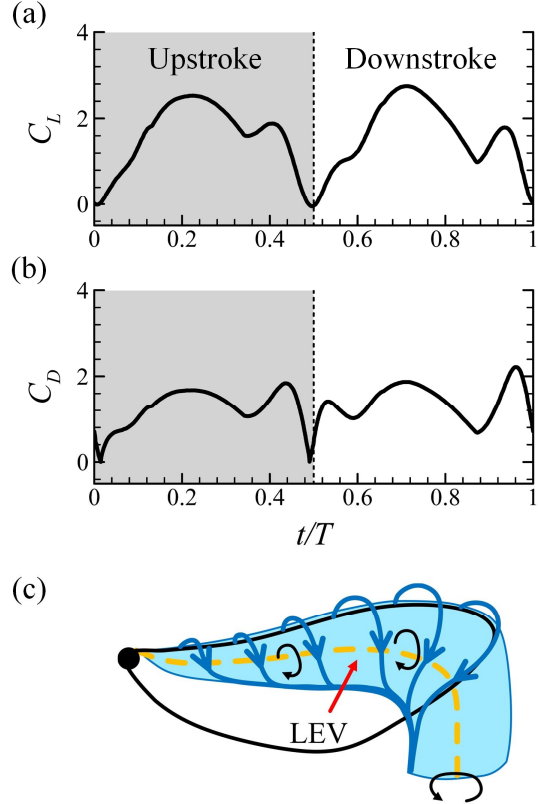

FIG. 3. Lift (a) and drag (b) coefficients of a dronefly in a flapping period, adopted from Liu and Sun (2008). (c) A sketch of the LEV on a flapping wing.

The large lift peak in the mid-portion of the upstroke or downstroke, due to the delayed-stall mechanism, provides approximately 60%–70% of the stroke-cycle mean lift (Liu and Sun 2008; Lee, Choi and Kim 2015). We thus see that insects use the lift of the wings to provide the weight-supporting vertical force and that the high lift is mainly produced by the LEV that attaches to the wing, i.e. by the delayed-stall mechanism.

The above results for medium and large insects, however, do not apply to the miniature insects, because of the very large effect of air viscosity. The viscous effect on the flow around an insect wing depends on the Reynolds number ($Re$), which is the ratio of inertial to viscous forces acting on a volume of air. The smaller the $Re$, the larger the viscous effect. $Re$ is proportional both to the chord length and the relative velocity of the wing (see Sec. II). As insect size becomes smaller, both these two quantities decrease. Therefore, *Re* decreases rapidly with the decreasing insect-size. For the miniature insects such as the tiny wasp *Encarsia formosa* (the wing length is approximately 0.6 mm), *Re* is approximately 10 (Weis-Fogh 1973). At such a low *Re*, the viscous effect of the air is so large that a miniature insect moves through the air as would a bumble bee move through mineral oil. The LEV would be significantly defused owing to the very large viscous effect and little lift could be generated (Miller and Peskin, 2004; Lyu, Zhu and Sun, 2019). Therefore, Miniature insects must use different wing kinematics and aerodynamic mechanisms from those of the medium and large insects.

In the last ten more years, much work has been done in the study of the mechanics of flight in miniature insects: novel flapping modes have been discovered and new mechanisms of aerodynamic force generation have been revealed; progress has also been made on the fluid-mechanics related flight problems, such as flight power requirements and flight dynamic stability. In this Colloquium article we attempt to review the significant works done so far in the area of miniature insect flight and discusses potential future directions. Furthermore, it provides the background necessary to do research in the area.

## II. GOVERNING EQUATIONS

The governing equations of the flow around an insect are the incompressible Navier–Stokes equations:

$$\nabla \cdot \mathbf{u} = 0 \tag{1}$$

$$\frac{\partial \mathbf{u}}{\partial t} + \mathbf{u} \cdot \nabla \mathbf{u} = -\frac{1}{\rho}\nabla p + \nu \nabla^2 \mathbf{u} \tag{2}$$

where $\mathbf{u}$ is the fluid velocity, $p$ the pressure, $\rho$ the density, $\nu$ the kinematic viscosity, $\nabla$ the gradient operator and $\nabla^2$ the Laplacian operator.

Using the wing chord length $c$ as reference length, the mean flapping speed $U$ as reference speed ($U=2\Phi r_2 f$, where $\Phi$ is flapping amplitude, $r_2$ the radius of gyration of wing and $f$ the flapping frequency) and the flapping period $T$ as the reference time ($T=1/f$), the non-dimensionalized Navier-Stokes equations can be written as:

$$St\frac{\partial \mathbf{u}^*}{\partial t^*} + \mathbf{u}^* \cdot \nabla \mathbf{u}^* = -\nabla p^* + \frac{1}{Re}\nabla^2 \mathbf{u}^* \tag{3}$$

$$\nabla \mathbf{u}^* = 0 \tag{4}$$

where the symbol "*" represents a non-dimensional quantity: $\boldsymbol{u}^*=\boldsymbol{u}/U$; $p^*=p/(\rho U^2)$; and $Re=cU/\nu$ (Reynolds number); and $St=(c/U)/(T/2)$ (Strouhal number). It is well known (see e.g. Batchelor, 1970) that $Re$ represents the ratio of the inertial force (the part of inertial force caused by convection acceleration) to the viscous force of the fluid, and that $St$ represents the ratio of the characteristic time of body motion (here, $T$) to the characteristic time of fluid convection ($U/c$). Since both $U$ and $T$ are related to flapping frequency $f$, $St$ can be written as $St = c/(\Phi r_2)$. For many insects, $\Phi$ is approximately 120° and $c/r_2$ is approximately 1.65 (Weis-Fogh, 1973), hence $St \approx 0.3$. $Re$ is proportional to both $c$ and $r_2$. When the size of an insect is very small, both $c$ and $r_2$ are very small, then $Re$ will becomes very low. That is, for a miniature insect, the viscous-force term in Eq. (3) (the second term on the right) will be very large, i.e. the viscous effect will be very strong.

When the wing motion of an insect is given (usually by measurement), Eqs. (3) and (4) can be numerically solved to give the flows around and the aerodynamic force acting on an insect. The equations are also used in the analysis of flows and aerodynamic mechanisms.

### III. A TYPICAL MINIATURE INSECT: *ENCARSIA FORMOSA*

As aforementioned (Sec. I), in order to generate aerodynamic forces necessary to fly, miniature insects must have different flapping kinematics and aerodynamic mechanisms from those of medium and large insects. What flapping pattern the miniature insects use and how they generate the necessary aerodynamic forces? Cheng and Sun (2018) took on these questions by studying the hovering flight of a typical miniature insect, tiny wasp *Encarsia formosa*; its wing length ($R$) is approximately 0.6 mm and its mass ($m$) about 0.02 mg. They first used high-speed cameras to measure the detailed wing kinematics, and then, based on the wing motion data, they solved the Navier-Stokes equations and obtained the flows around and the aerodynamic forces on the insect.

They found that the miniature wasp has a distinctive pattern of wing motion very different from that of medium and large insects. Figure 4 gives the wing motion diagram: Let $T$ be the period of the flapping cycle. A flapping cycle can be divided into four phases. First, in the upstroke, the wings accelerate rapidly downward and backward at almost 90° angle of attack [Fig. 4(a); $t/T \approx 0\text{-}0.25$; the first phase], and then they close up on the right and left sides at the back of the insect, while at the same time moving slowly upwards at almost 0° angle of attack [Fig. 4(a), $t/T \approx$ 0.25-0.5; the second phase]. Then, in the beginning of the downstroke, the wings quickly rotate about their trailing edges [Fig. 4(b), $t/T \approx$ 0.5-0.6; the third phase] and then sweep forward at high angle of attack [Fig. 4(a), $t/T \approx$ 0.6-1.0; the fourth phase]. The wing motion in the first phase, fast accelerating downward and backward at almost 90° angle of attack, resembles that of the stroking oars of a boat; this phase is referred to as 'impulsive rowing'. The second phase is referred to as 'clap and slowly moving up'. The third phase is the well known 'fling' motion, first discovered by Weis-Fogh (1973) in this species (more discussion on this motion will be made later in this section and in Sec. V). The fourth phase is referred to as forward sweep. As mentioned in Sec. I, medium and large insects flap their wings approximately in a plane, or they have planar upstroke and downstroke. For the tiny wasp, the upstroke has a very deep U-shape [Fig. 4(a)], and the downstroke also has a U-shape, but it is much shallower; in an entire wingbeat cycle, the wing tip follow a 'twisted' figure-of-eight loop [Fig. 4(a)].

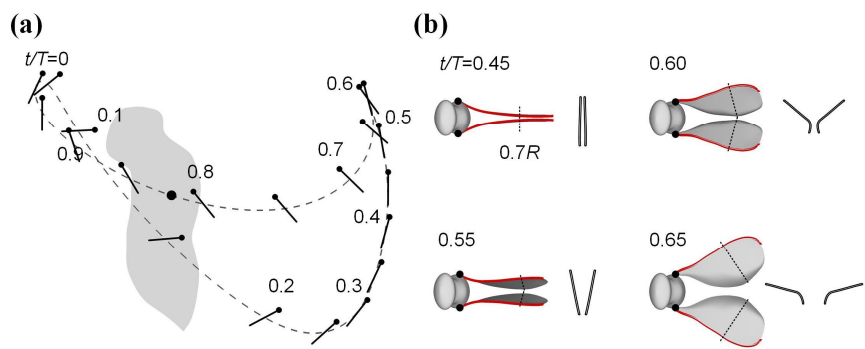

FIG. 4. (*a*) Stroke diagram showing the wing motion of *E. formosa*. The dashed curve indicates the wing-tip trajectory (projected onto the symmetrical plane of the insect); black lines indicate the orientation of the wing at 20 temporally equidistant points, with dots marking the leading edge; the black dot defines the wing-root location on the insect body. (*b*) Wing motion at dorsal stroke reversal, showing the 'fling'. *R*, wing length. Adapted from Cheng and Sun (2018).

What do these odd features of wing flapping give in terms of aerodynamic force production? The computed aerodynamic forces are shown in Fig. 5; in the figure and in the rest of this paper, the vertical and horizontal components of the total aerodynamic force of a wing are referred to as vertical force and horizontal force, respectively. When the wing does not translate in horizontal plane, the vertical force is not the lift of the wing and the horizontal force is not the drag of the wing. The velocity at the radius of gyration of the wing is used to represent the velocity of the wing [Fig. 5(e)]. Lift and drag are defined as the components of the total aerodynamic force that are perpendicular and parallel to the velocity of the wing, respectively. As seen from the figure, a large vertical-force peak is produced during the impulsive rowing [Fig. 5(a), $t/T \approx 0$–0.2] and another smaller one during the 'fling' and the beginning of the forward sweeping [Fig. 5(a), $t/T \approx 0.55$–0.75]. The other parts of the cycle do not produce any positive vertical force; during the 'clap and slowly moving up' motion, the motion is slow and drag is small, and moreover, two wings 'clap' and move somewhat as one wing, further reducing their drag or the negative vertical force (Cheng and Sun, 2019). The impulsive rowing contributes 70% of the total vertical force, and the 'fling' and the beginning of the forward sweeping contribute the other 30%.

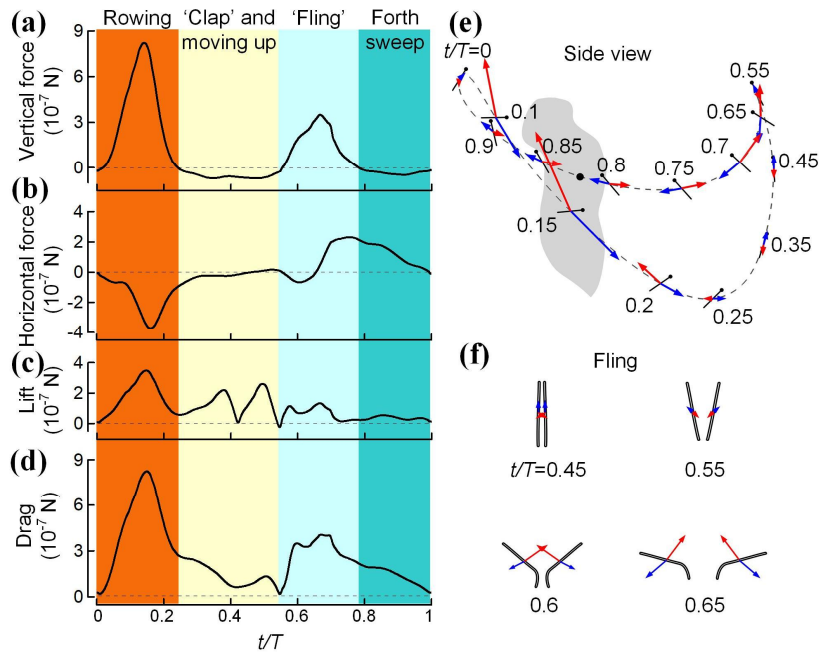

FIG. 5. Aerodynamic forces. (*a–d*) Vertical force, horizontal force, lift and drag of a wing. (*e*) Diagram showing the wing position, velocity vector and force vector at various times in one stroke cycle (side view). (*f*) Back view of the wing in 'fling' motion, represented by the wing section at 0.7*R*, and the corresponding velocity vector and force vector. Blue arrow, velocity; red arrow, force. Adapted from Cheng and Sun (2018).

The authors further explained that the large vertical-force peak during the impulsive rowing is mainly from the drag of the wing (in this period, the wing surface is almost horizontal) and that the large drag is due to the fast acceleration of wing (referred to as 'impulsive rowing' mechanism). As for the vertical-force peak during the 'fling' and the beginning of the forward sweeping, it is also mainly from the large drag of the wing [see Fig. 5(a), (d) and (f)]. The production mechanism of the large drag is that the opening of the wing pair generates a low-pressure region of rapidly swirling air which persists to the beginning of the forward sweeping, creating the large drag (referred to as 'fling' mechanism). The 'fling' mechanism was discovered by Weis-Fogh (1973), Maxworthy (1979) and Spedding and Maxworthy (1986); it was further studied by many researchers, e.g. Sun and Yu (2003), Miller and Peskin (2005), Lehman and Pick (2007), Arora *et al.* (2014) and Santhanakrishnan (2014).

We thus see that the tiny wasp uses the drag produced by distinctive wing motion and novel aerodynamic mechanisms ('impulsive rowing' and 'fling') to offset the otherwise formidable effects of increased viscosity. These two mechanisms can also be explained by examining the flow equations, Eq. (3), re-written here as:

$$\underbrace{St\frac{\partial \boldsymbol{u}^*}{\partial t^*}}_{\text{inertial force (local rate of change)}} + \underbrace{\boldsymbol{u}^* \cdot \nabla \boldsymbol{u}^*}_{\text{inertial force (convection)}} = -\nabla p^* + \underbrace{\frac{1}{Re}\nabla^2 \boldsymbol{u}^*}_{\text{viscous force}} \qquad (5)$$

When *Re* is very small, the viscous effect, represented by the second term in the right of Eq. (5) is very large. As mentioned in Sec. II, the characteristic time used in *St* is usually *T* (the flapping period). But when the wing has very large velocity change (i.e. large acceleration) in a short time, say 0.1*T*, there will be a new, smaller time scale. Using this smaller characteristic time as reference time, *St* would be one order of magnitude larger. This would make the inertial force, the first term in the left of Eq. (5), large, overcoming the large viscous effect.

That miniature insects may use drag mechanism for flight was suggested before (Horridge, 1956; Jones *et al.* 2015). Two modes of wing motion that can use drag to produce weight-supporting vertical force (Fig. 6) were suggested by Jones *et al.* (2015). The downward motion of wing in Fig. 6(a) looks similar to the rowing motion of the tiny wasp, but there is a major difference between the two: the wing in Jones *et al.*'s suggested motion moves at approximately constant speed, while the real miniature insect moves with fast acceleration. Without the large acceleration, the required large drag (vertical force) cannot be produced.

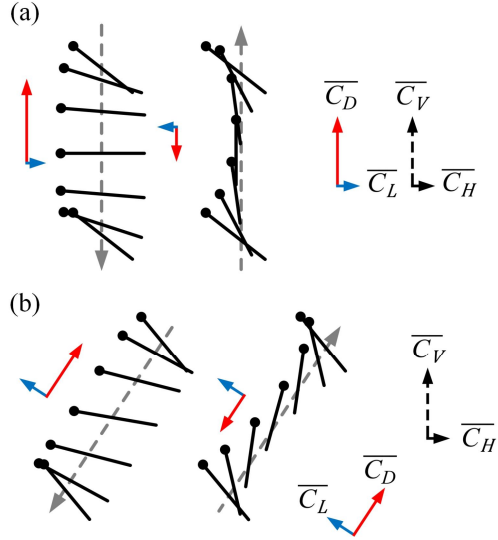

FIG. 6. Idealized wing kinematics (represented by a sections of the wing). The dot represents the leading edge of the wing, and the down- and upstrokes are shown separately. The solid red arrows indicate the direction and relative magnitude of dimensionless drag ($C_D$) during each down- or upstroke, and the solid blue arrows indicate the direction and relative magnitude of dimensionless lift ($C_L$). The solid axes on the right indicate the net $C_D$ and $C_L$ during a flapping cycle. The direction of dimensionless vertical ($C_V$) and dimensionless horizontal ($C_H$) force always face the same direction with respect to the global frame (dashed arrows). (a) A vertical, drag-based stroke uses only $C_D$ to produce CV. (b) A tilted, hybrid stroke   uses both $C_L$ and $C_D$ to produce $C_V$. Modified from Jones *et al*. (2015).

## IV. A SERIES OF MINIATURE INSECTS OF DIFFERENT SIZE
### A.  Some miniature insects of different size

As aforementioned, generally the up- and downstrokes of medium and large insects are planar (Fig. 2). But for the tiny wasp *E. formosa* discussed in the previous section (Sec. III), the planar upstroke commonly used by medium and large insects changes to a deep U-shape upstroke [Fig. 4(a)]. The first part of the deep U-shape upstroke (wings fast accelerating downward) produces a very large vertical-force by the 'impulsively rowing' mechanism and the second part it (wings clapped and slowly moving up) produces very small negative vertical-force; thus the deep U-shape upstroke gives a large mean vertical force and overcoming the otherwise formidable effects of increased viscosity.

The tiny wasp *E. formosa* is a very small insect, $R \approx 0.6$ mm and $Re \approx 10$ ($m \approx 0.02$ mg). Lyu, Zhu and Sun (2019) examined the literature and noticed that small fruitflies, a relatively large miniature insect, whose $R \approx 3$ mm and $Re \approx 80$ ($m \approx 0.72$ mg), have a shallow U-shape upstroke (Fry, Sayaman and Dickinson, 2005; Meng and Sun, 2015). The wing-tip trajectory of the fruitfly, compared with that of the tiny wasp, are plotted in Fig. 7: the U-upstroke of the fruitfly (at the top of Fig. 7) is much shallower than that of the tiny wasp (at bottom of Fig. 7). This inspired Lyu, Zhu and Sun (2019) to conjecture that as the insect size decreases, i.e. $Re$ decreases, deeper and deeper U-shape upstroke would be used to overcome the larger and larger viscous effects, as shown by the wing-tip trajectories drawn using doted lines in Fig. 7.

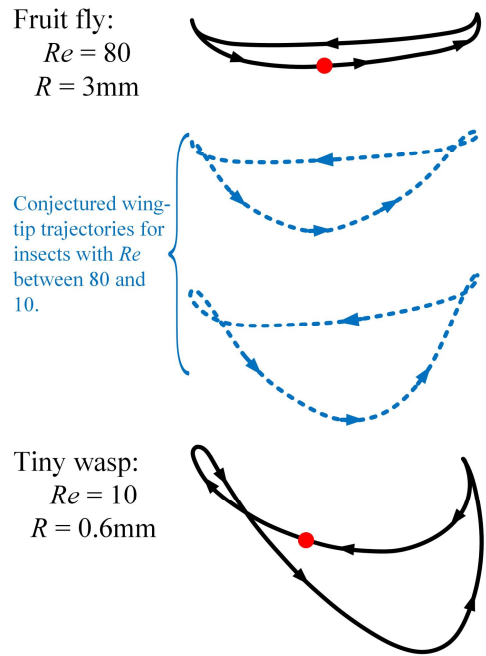

FIG. 7. Top, the wing-tip trajectory (projected onto the symmetrical plane of the insect) of a fruitfly ($R$=3 mm), which is a relatively large miniature insect. Bottom, the wing-tip trajectory of a tiny wasp ($R$=0.5 mm), which is a typical miniature insect. Middle, conjectured wing-tip trajectories for insects with size in between.

They measured the wing motions of several insects with size smaller than that of the small fruitfly and larger than that of the tiny wasps. These insects are: vegetable leafminer *Liriomyza sativae* (LS) ($Re≈40$, $R≈1.5$ mm and $m≈0.25$ mg), biting midge *Forcipomia gloriose* (FG) ($Re≈29.8$, $R≈1.4$ mm and $m≈0.2$ mg), biting midge *Dasyhelea flaviventris* (DF) ($Re≈23.7$, $R≈0.95$ mm and $m≈0.08$ mg), gall-midges *Anbremia sp.* (AS) ($Re≈17.4$, $R≈1.3$ mm and $m≈0.05$ mg) and thrips *Frankliniella occidentalis* (FO) ($Re≈13.5$, $R≈0.79$ mm and $m≈0.02$ mg). The stroke diagrams showing the flapping mode of the these insects are plotted in Fig. 8 (those of small fruitfly *D. virilis* and tiny wasp *E. formosa* are also included in the figure). As aforementioned, medium and large insects, for example dronefly (Fig. 2), has approximately planar upstroke and downstroke, while from Fig. 8, we see that for the miniature insects, the upstroke changes to be U-shaped, and as size decreases, deeper and deeper U-shape upstroke are employed [Figs. 8(a)–(g)]. This is just what the authors has conjectured (see Fig. 7).

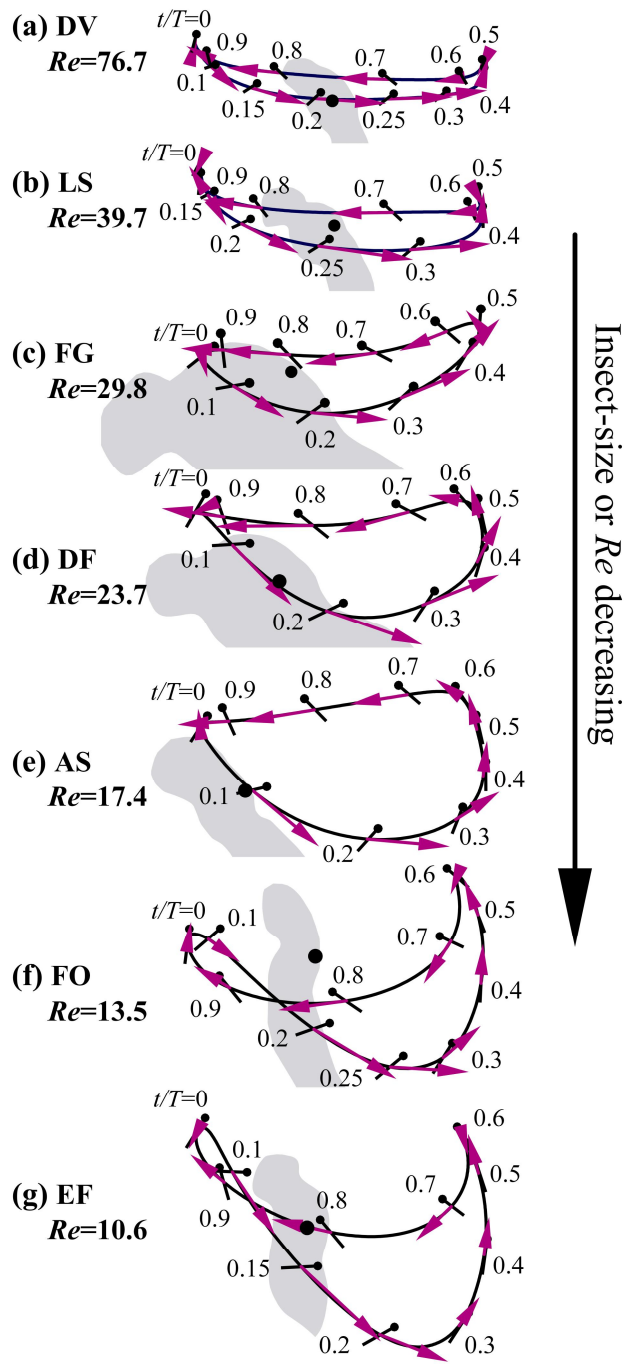

FIG. 8. Stroke diagrams showing the wing motions of the insects. The solid curve indicates the wing-tip trajectory ((projected onto the symmetrical plane of the insect); black lines indicate the orientation of the wing at various times in one stroke cycle, with dots marking the leading edge; a black dot defifines the wing-root location on the insect body; the purple arrow represents the velocity of the wing at the radius of gyration. Adapted from Lyu, Zhu and Sun (2019).

Moreover, for the two smallest of these insects, thrip *F. occidentalis* and tiny wasp *E. formosa*, the downstroke also becomes U-shaped, but the downstroke U-shape curve is shallower than that of the upstroke [Figs. 8(f) and 8(g)]; thus in a entire wingbeat cycle, the wing tip follow a 'twisted' figure-of-eight loop.

Using the wing motion data, the authors (Lyu, Zhu and Sun, 2019) solved the Navier-Stokes equations [Eqs. (3) and (4)] and obtained the aerodynamic forces on the insects, and they showed the following. For the relatively-large miniature insects, the U-shape upstroke produces a larger vertical force than planar upstroke by having a larger wing velocity. For the very small ones, similar to the tiny wasp *E. formosa* discussed in Sec. III, the U-shape upstroke becomes very deep, and in its first first phase, the wing smashes on the air ('impulsive rowing') and generates a very large drag directing upwards (vertical force), whereas in its second phase, the wing slices through the air slowly and generates a very small drag directing downward. They also showed that the 'fling' mechanism were also used by some of the miniature insects.

**B. An even smaller tiny insect (a tiny beetle with bristled wings)**

In the above sub-section, the flight of some miniature insects of different size were discussed, and the smallest ones are the tiny wasp *E. formosa* and the thrip *F. occidentalis* ($m \approx 0.02$ mg). But there are many even smaller winged insects, and most of these smallest winged insects have bristled wings (Polilov, 2005; Farisenkov *et al.*, 2020); a bristled wing is sketched in Fig. 9.

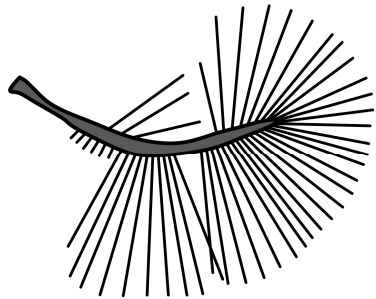

FIG. 9. A sketch of the bristled wing of a tiny beetle *Paratuposa placentis*, plotted according to the photograph by Farisenkov *et al*. (2022).

Do these even smaller insects use an even deeper U-upstroke, or they use both deep U-upstroke and deep U-downstroke, or they use other novel flapping mode? Recently Farisenkov *et al.* (2022) studied the flight of a tiny beetle *Paratuposa placentis* whose $m$ is only 0.0024 mg ($R$=0.49 mm, $Re$=9). They used high-speed videography to obtain the wing kinematics, electron microscopy measurement to obtain the morphological data, and the method of computational fluid mechanics to compute the flow and aerodynamic forces.

The measured wing kinematics is shown in Fig. 10 by a diagram of wing-tip trajectory (side view) and wing orientation at 20 points of equal time interval. In Fig. 10, curve ABC is the wing-tip trajectory of the upstroke and curve CDA is that of the downstroke. An upstroke (or downstroke) can be divided into two phases, called as power phase and recovery phase, respectively, by Farisenkov *et al.* (2022).

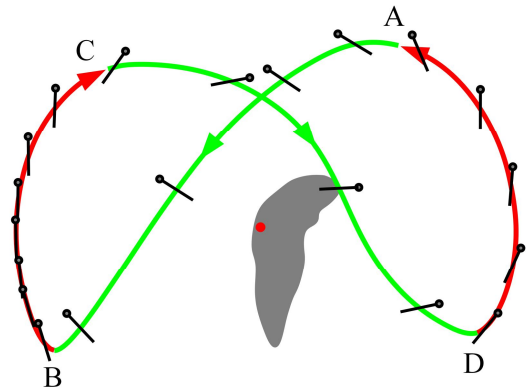

FIG. 10. Stroke diagrams showing the wing motions of *Paratuposa placentis*. The solid curve indicates the wing-tip trajectory (projected onto the symmetrical plane of the insect); black lines indicate the orientation of the wing at various times in one stroke cycle, with dots marking the leading edge; a red dot defines the wing-root location on the insect body. Adapted from Farisenkov *et al*. (2022).

First, let us look at the upstroke. In the first phase (power phase) of the stroke (Fig. 10, from A to B, colored green), the wings move downward and backward at large velocity and acceleration at nearly 90° angle of attack. In the second phase (recovery phase) of the upstroke (Fig. 10, from B to C, colored red), the wings move upwards slowly at nearly zero angle of attack [at the beginning of the moving up, the two wings close up ('clap') at the back of the insect (this can be seen from Fig. 11, which is to be discussed below)], and near the end of the moving up, the wings start to separate from each other. Note that the upstroke of the tiny beetle is similar to the deep U-shape upstroke of the tiny insects discussed in above sections (Sec. III and Sec. IV-A). Next, we look at the downstroke. In the first phase of stroke (Fig. 10, from C to D, colored green), similar to the first part of the upstroke, the wings move at large velocity and acceleration at nearly 90° angle of attack, but here they move downward and forward, not downward and backward. In the second phase of the stroke (Fig. 10, from D to A, colored red), similar to the case of the upstroke, the wings move upward slowly at nearly zero angle of attack; the two wings are also closed, but now the wings are in the front of the insect. The downstroke is also similar to a deep U-shape stroke. Since the both the upstroke and downstroke are deep U-shaped, the wing-tip trajectory is a figure-of-eight loop (Fig. 10).

On the basis of the measured kinematic and morphological data of the tiny beetle, the authors computed the flow around the insect and obtained the aerodynamic force on the bristled wings, as shown in the sketch in Fig. 11. As seen from Fig. 11, very large, nearly upward pointing, drag is produced by the wing in the first phase (power phase) of the U-shape upstroke when the wings moving at large velocity and acceleration at nearly 90° angle of attack ('impulsive rowing' mechanism) and small, nearly upward pointing, drag is produced in the second phase (recovery phase) in which the wings move upward slowly. So is the case with the downstroke (Fig. 11).

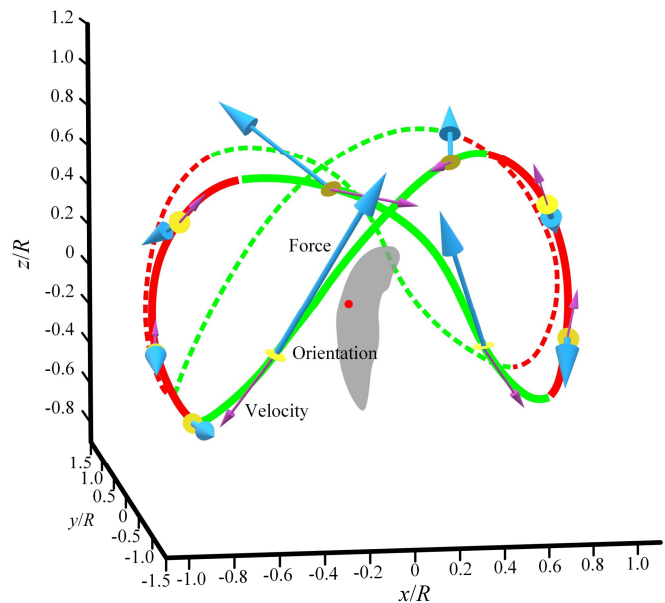

FIG. 11. Wing-tip trajectories and direction of aerodynamic force and wing velocity: Cyan arrows show aerodynamic force; magenta arrows show wing-tip velocity; yellow discs show dorsal surface orientation of the wing at nine time instants. Opaque and transparent curves correspond to right and left wing, respectively. Adapted from Farisenkov *et al*. (2022).

The smallest insects discussed in earlier sections (Sec. III and Sec. IV-B), i.e. tiny wasp *E. formosa* and thrip *F. occidentalis*, mainly use a deep U-upstroke (impulsive rowing mechanism) and a shallower U-downstroke to overcome the strong viscous effect. The tiny beetle *P. placentis* discussed in this sub-section is even smaller. From the above discussion, we see that the wing motion style of the tiny beetle *P. placentis* is only a little different from that of the thrip *F. occidentalis* and tiny wasp *E. formosa*: Comparing Fig. 8(f) and (g) with Fig. 10, it is seen that they all have a deep U-shape upstroke; the only difference is that the tiny beetle *P. placentis* also has a deep U-shape downstroke while the thrip *F. occidentalis* and tiny wasp *E. formosa* have only a shallower U-shape downstroke. For the thrip *F. occidentalis* and tiny wasp *E. formosa*, in an entire wingbeat cycle, the wing tip follow a 'twisted' figure-of-eight loop [Fig. 8(f) and (g)], while for the tiny beetle *P. placentis*, the wing tip follow a 'nomal' figure-of-eight loop (Fig. 10).

In Sec. III and the present section (Sec. IV), we have discussed eight species of miniature insects whose wing kinematics have so far been measured. The size of these insects, represented by body mass $m$, ranges from $m \approx 0.72$ mg (small fruitfly *D. virilis*) to $m \approx 0.0024$ mg (tiny beetle *P. placentis*). From the measured data (Fig. 8 and Fig. 10), the following trend of variation in flapping pattern is observed: The planar upstroke commonly used by the medium and large insects changes to a U-shaped upstroke, and as the size of insect becomes smaller, deeper and deeper U-shaped upstroke is used. When $m$ decreases to about 0.02 mg (tiny wasp *E. formosa* and thrip *F. occidentalis*), the downstroke also becomes U-shaped, but the U-shape is shallower [(Fig. 8 (f) and (g)]. When $m$ decreases to about 0.0024 mg (tiny beetle *P. placentis*), the downstroke also becomes deep U-shaped (Fig. 10). The features of a deep U-shape up- or downstroke are as follows: in its first phase [wings fast accelerating downward and backward (or forward)], a very large vertical-force is produced by the 'impulsively rowing' mechanism and in its second phase (wings clapped and slowly moving up), only very small negative vertical-force is produced, thus the deep U-shape up- or downstroke gives a large mean vertical force.

## V. BRISTLED VERSUS MEMBRANOUS WINGS

Many of the smallest flying insects have bristled wings (Fig. 9), e.g. thrips (Lewis, 1973), tiny beetles (Polilov, 2005) and fairyflies (Huber and Noyes, 2013). Some of them have partially bristled wings (wing consists of a membranous part and a brim of marginal bristles), e.g. tiny wasp *E. formosa* (Weis-Fogh, 1973). For the tiny beetle *P. placentis*, it has recently been shown that using the bristled wings is a prerequisite for its fast flight (Farisenkov *et al.*, 2022). Therefore, it is of interest to have some discussion on bristled wings. Here we consider following questions: whether or not a bristled wing can generate as much aerodynamic force as a membranous wing, how aerodynamic force is generated by the bristles, what advantages a bristled wing has over a membranous wing, and are the bristles, which are very slender, stiff enough to withstand the aerodynamic force?

**A. Aerodynamic force**

Sunada *et al.* (2002) measured the aerodynamic forces on a model bristled wing and a solid-plate wing of the same shape. Different simple wing motions were considered: azimuthally rotating at a constant angular speed and at a constant angular acceleration and translating at a constant speed and at a constant acceleration. They showed that at *Re* around 10 and below, the aerodynamic forces acting on the model bristled wing were only a little smaller than those on the solid-plate wing [miniature insects with bristled wings fly at *Re*≈10 and below (see, e.g. Sunada *et al.*, 2002; Lyu, Zhu and Sun, 2019; Farisenkov *et al.*, 2022)]. Similar experimental investigation (Lee and Kim, 2017) showed that at higher *Re*, a bristled wing produced much less aerodynamic force than the corresponding solid-plate wing. In the above experiments (Sunada *et al.*, 2002; Lee and Kim, 2017), the wing model has a rectangular planform and the bristles are in the chordwise direction. In a real bristled wing (as sketched in Fig. 9), in the inner part of the wing, the bristles point forward and backward, while in the outer part of the wing, the bristles point laterally. Kolomenskiy *et al.* (2020) performed experimental and numerical studies using a model bristled wing of realistic morphology of a miniature beetle. It was found that in the considered biologically relevant regimes of flow parameters, the bristled wing produced between 60% to 96% of the aerodynamic force of an equivalent membranous wing. In the experiment, the wing performed constant speed rotation, not real flapping motion. In some recent numerical studies, realistic wing motions and realistic bristled wing morphology of a tiny beetle (Farisenkov *et al.*, 2022) and a tiny wasp (Jiang *et al.* 2022) were used. It was shown that the bristled produced aerodynamic force more than 80% of that of the corresponding membranous wing.

The above studies show that at *Re* of the bristled winged miniature insects (*Re*≈10 and below), the bristled wing can produce aerodynamic forces only a little smaller than the corresponding membranous wing.

**B. Production mechanisms of aerodynamic force**

A bristled wing is morphologically very different from a membranous wing of the same outline. The membranous wing is solid, like a flat plate, while the bristled wing has gaps between the bristles, like a comb (Fig. 9); the gap width is approximately 10 times of the bristle diameter (Sunada *et al.*, 2002; Jones *et al.*, 2016). Their aerodynamic force production mechanisms should be different. As discussed above (Sec. III and Sec. IV), the wings of the smallest insects move at nearly 90° angle of attack, producing large drag to provide the weight-supporting vertical force. For a flat-plate wing moving at nearly 90° angle of attack, it is obvious that the drag mainly come from the pressure difference between the windward and the rearward surfaces of wing (frictional force is tangential to the surfaces and has little contribution to the drag). For a bristled wing, the situation is different. Each bristle of the wing is a very slender 'cylinder'. When the Reynolds number of the wing is 10 (based on the mean chord length of the wing), the Reynolds number of the bristle is only about 0.05 (based on the diameter of the bristle). That is, the flows around the bristles must be Stokes flows and the drag on the bristles must be similar to that of a cylinder in Stokes flow, half contribution by friction force and half by pressure force. Barta and Weihs (2006) modeled the bristled wing by a row of parallel slender bodies (rod-like ellipsoids of slenderness ratio smaller than 0.01), and solved the Stokes equation analytically; in this model, obviously the bristles generate forces by Stokes-flow (or creeping flow) mechanism. Since they solved the Stokes equation, the Reynold number of their model wing (*Re* based on wing chord length) needs to tend to zero, i.e., $Re \ll 1$; the flow near the 'bristle' (near field) and the flows in the far field are all Stokes flows. Lee, Lee and Kim (2020), Lee and Kim (2021) and Wu and Liu and Sun (2021) modeled the bristled wing by a row of two-dimensional circular cylinders and solved the Navier-Stokes equation numerically. Here *Re* can be any value as long as it is low enough for turbulence not occurring (for miniature insects, *Re*≈10, this is of course not a problem). Lee, Lee and Kim (2020) considered the effects of varying the gap width between bristles and varying *Re* on the flows around the bristles. Lee and Kim (2021) and Wu, Liu and Sun (2021) investigated the effects of intermittent gusty and wing acceleration, respectively. Both group's results showed the following. When *Re* of the bristled wing is about 10 or less (the Reynolds number of a bristle is about 0.05 or less), the flow near each bristle is Stokes flow in nature: the streamlines in the front of the bristle and those in the back are symmetrical; the surface frictional force and pressure force have approximately the same contribution to the drag on the bristle. The flow away from the bristles (far field) resembles that of the corresponding flat-plate wing. Recently, Jiang *et al.* (2022) used realistic configuration of the bristled wings of a tiny wasp and realistic flapping motion in their numerical solution of the Navier-Stokes equations. They also found that the aerodynamic force on the bristle is approximately half contributed by surface pressure force and half by the surface friction force.

From above results, we see that the drag production mechanism of the bristled wing is different from that of the membranous wing: For the membranous wing, the flow is blocked by the wing, giving a positive pressure on the windward surface and a negative pressure on the leeward surface; the drag is due to the pressure forces (the frictional stress has almost no contribution). For the bristled wing, each bristle operates in a creeping flow and produces thick and strong shear layers; strong viscous force generates a very large pressure difference between the windward and leeward surfaces of each bristle and very large frictional stress on the bristle surface, resulting in a large drag on each bristle, and the drag is equally contributed by the pressure and frictional forces.

## C. Deformation of the bristles

As discussed above (sub-section A), a bristled wing operating in the biologically relevant Reynolds number regimes (*Re*≈10 and below, based on the wing chord length) produces an aerodynamic force close to that of a membranous wing. The projected wing area of a bristled wing is much smaller than that of the equivalent membranous wing; for example, the projected wing area of the bristled wing of the miniature beetle *P. placentis* is only 15% of that of the equivalent membranous wing (Kolomenskiy *et al.*, 2020). This means that the force on per unit area of a bristled wing is much larger than that of the corresponding membranous wing. Furthermore, the bristles are very slender and seem to be flexible. One might intuitively think, whether or not the bristles would have large deformation and if the wing geometry could be maintained.

Jiang *et al.* (2022) recently investigated this problem. They considered the bristled wings of parasitoid wasp *Anagrus Haliday*. They first measured the morphological characteristics and the Young's modulus of the bristles, and then performed fluid-structure-interaction computations [i.e. solving the Navier-Stokes equations, Eqs. (3) and (4), coupled with the equilibrium equation for elastic body] of a model bristled wing. Their measurements showed that the bristles have a conical tubular structure, tapering toward the tip, and that the Young's modulus of the bristles is in the range of 18.6-25.2 GPa, which is much higher than that conventionally considered value for bristles, 8-10 GPa (Seale *et al.*, 2018). Their computations showed that at extreme flow velocity and angle of attack (90°) of tiny wasps given in literature, the bristles only deflect marginally (a bending angle of less than 0.3°). That is, the bristled wing at high aerodynamic loading has negligible deformation and wings geometry could be maintained.

The above results are for the parasitoid wasp *A. Haliday*. For other miniature insects with bristled wings, no such detailed quantitative study has been made yet. However, from high-speed video recordings of several free-flying miniature insects with bristled wings, e.g. thrips in take-off flight (Santhanakrishnan *et al.*, 2014) and hovering flight (Lyu, Zhu and Sun, 2019), tiny wasp *E. formosa* in hovering and forward flight (Cheng and Sun, 2018, 2021) and tiny beetle *P. placentis* in hovering and forward flight (Farisenkov, 2022), no noticeable wing (bristles) deformation are observed. This indicates that the slender bristles have special morphological structure and material characteristics for resisting deformation.

**D. Merits of the bristled wings**

As seen above (sub-section A), a bristled wing operating at *Re* of the order of 10 or less could produce a aerodynamic force close to that of a membranous wing. And it has been observed that the wings of many smallest insects are bristled rather than membranous. Then, one would ask the question: what merits a bristled wing possesses, compared with a membranous wing? Researchers have done some works regarding this question. Here, we discuss some results of these works.

The wing surface area of a bristled wing is much smaller than that of the membranous wing of the same outline. Therefore, it has been commonly considered that a bristled wing is much lighter than the equivalent membranous wing, and hence the inertial power for flapping a bristled wing would be much less than that for the membranous wing (see e.g. Sunada, Kawachi and Yasuda, 2003; Weihs and Barta, 2008). But only recently, a quantitative study on this possible merit has been conducted. In their study of the tiny beetle *P. placentis*, Farisenkov *et al.* (2022) determined the mass and moment of inertia of the bristled wing on the basis of measured data. They estimated the mass and moment of inertia of the equivalent membranous wing by using the wing thickness of some of larger membranous-winged insects (body length about 1 mm). They found that the bristled wings are lighter than the equivalent membranous wings and have smaller moment of inertia. With the moment of inertia of the bristled and membranous wing evaluated, they numerically solved the Navier-Stokes equations and computed the inertial and aerodynamic powers of the free flying tiny beetle [the power required for flight is the sum of the aerodynamic and inertial power (see Sec. VI)]. They found that with the bristled wing, the instantaneous power may reach up to 110 W per kg body mass (W·kg$^{-1}$) in the power phase of the flapping cycle, while with the membranous wings, the value can reach up to 180-210 W·kg$^{-1}$. This shows that with bristled wings, the tiny beetle needs much less power than the case of with membranous wings. (It should be noted that in the study, the mass of the equivalent membranous wing, estimated by using the wing thickness of some of larger membranous-winged insects, might be too large. And furthermore, in the estimating the moment of inertia of the equivalent membranous wing, a uniform wing thickness was used, this may over estimate the moment of inertia. Future studies in miniature bristled and membranous winged insects of similar size might clear this point.)

Some bristled-winged miniature insects perform the 'fling' motion at the beginning of the downstroke (in the fling motion, the left and right wings are initially parallel and very close to each other, i.e. 'clapped', and then the wings rotate quickly around the trailing edge and open to form a V-shape). In the later stage of the fling motion, a relatively large vertical force can be produced; this aerodynamic force production mechanism is referred to as fling mechanism (Maxworthy, 1979). The fling motion can also enhance the vertical force production in the subsequent downstroke (see, e.g. Miller and Peskin, 2004; Cheng and Sun, 2021). The fling motion was identified for tiny wasp *E. formosa* in 1973 (Weis-Fogh, 1973) and since then many studies were made on this motion (e.g. Spedding and Maxworthy, 1986; Sun and Yu, 2003; Miller and Peskin 2005, 2009; Kolomenskiy *et al.* 2010; Arora *et al.* 2014). Membranous wings were considered in these studies; flow structure and aerodynamic force production mechanism, effects of initial distance between the two wings, effects of wing flexibility, etc., were studied. One of the important results of these studies is that the cost of flinging is rather high: the drag required to 'open' the wings apart may be an order of magnitude large than the force require to move a single wing with the same motion. Is it like this for bristled wings? Santhanakrishnan *et al.* (2014) investigated this question by using two-dimensional (2D) porous flat-plates to simulate the bristled wings. They numerically solved the Navier-Stokes equations for the porous plates and the corresponding solid plates. They found that compared with the solid wings, the porous nature of the wings contributes largely to drag reduction. This result may indicate that bristled wings, compared with solid wings, reduce the drag required to fling the two wings apart. However, a porous plate is not a bristled wing; it was desired to study the flows of bristled wing directly. Jones *et al.* (2016) and Wu, Liu and Sun (2022) studied the flows and aerodynamic forces of 2D bristled wings and the corresponding solid wings by numerically solving the Navier-Stokes equations. Kasoju *et al.* (2018) and Kasoju and Santhanakrishnan (2021) studied the case of three-dimensional (3D) wings experimentally, using robotic bristled wing models. Both the 2D numerical study and the 3D experimental studies showed that in fling motion, bristled wings significantly decrease the drag required to fling the wings apart, compared with the case of solid wings.

From the above discussions, two advantages of bristled wings over membranous wings are observed: being lighter in mass, hence needing smaller inertial power for their flapping; much less force being needed to open the wings in the fling motion. There may be some other merits of bristled wings to be discover later.

## VI. POWER REQUIREMENTS AND FLIGHT STABILITY: SIZE EFFECTS
### A. Power requirements

The wings of a flying insect must produce vertical force to support its weight and thrust to propel its body moving through the air. When producing these forces, the flight muscles must do work to move the wings against the aerodynamic drag and to accelerate the wing mass. The power needed to overcome the aerodynamic force is referred to as aerodynamic power and that to overcome the wing inertial force is referred to as inertial power. The sum of these two is the mechanical power which the flight muscles must deliver. When the metabolic rate of the flight muscles is known, the mechanical power can give the mechanochemical efficiency of the muscles; and knowing how much contribution the inertial power makes to the mechanical power can tell us whether or not an elastic muscle-system is essential for insects (Ellington, 1984c; Dudley and Ellington, 1990b). Therefore, the study of the mechanical power requirement is of great importance for understanding the physiological and bio-mechanical mechanisms of insect flight.

The calculation of the inertial power is relatively simple. When the flapping kinematics and the wing mass and its distribution are measured, the wing rotational velocity and acceleration and the moments of inertia of wing can be obtained; and from these data, the inertial power can be straightforwardly computed. The calculation of the aerodynamic power involving flow computation, i.e. numerically solving the Navier-Stokes equations [Eqs. (3) and (4)], or force and moment measurement using dynamically scaled wing model. Conventionally the mechanical power is denoted by $P$ and its flapping-cycle-mean value denoted by $\bar{P}$. Because the wings have acceleration in some parts of the cycle and deceleration in other parts, the inertial power, hence the mechanical power $P$ may become negative in some parts of the cycle. How the negative power fits into the power budget depends on the elastic energy storage of the flight muscle. When calculating $\bar{P}$, researchers commonly consider two limiting cases, and the real $\bar{P}$ lies between these limits (Dudley, 2000). One limiting case is 0% elastic energy storage, in which the negative work is ignored in the power budget; the other limiting case is 100% elastic energy storage, in which the negative mechanical work can be completely stored in an elastic element and later released to do positive work. The cycle-mean mechanical power divided by the mass of the insect is referred to as the mass-specific power, denoted as $P^*$,

$$P^* = \bar{P}/m \tag{6}$$

where $m$ is the mass of the insect. Clearly, $P^*$ means power required for per unit mass (weight) of the insect. Since $\bar{P}$ has two limiting values, so does $P^*$. In the case of 0% elastic power storage, the mass-specific power is denoted as $(P^*)_1$ and in the case of 100% elastic power storage, as $(P^*)_2$.

Earlier studies on flight power requirements were for medium and large insects, for which the wing kinematic and morphological data were available (e.g. Dudley and Ellington, 1900a; Lehman and Dickinson, 1997; Sane and Dickinson, 2001, Sun and Tang, 2002b; Young *et al.*, 2009). In recent years, the required data have been measured for a number of miniature insects (Lyu, Zhu and Sun, 2019; Farisenkov *et al.*, 2022), and their power requirements can be computed. Lyu and Sun (2021) calculated the power requirement of six species of miniature insects in hovering flight; these insects are vegetable leafminers *L. sativae* (LS) ($m\approx 0.25$ mg), biting midges *F. gloriose* (FG) ($m\approx 0.2$ mg), biting midges *D. flaviventris* (DF) ($m\approx 0.08$ mg), gall midges *A.* sp. (AS) ($m\approx 0.05$ mg), thrips *F. occidentalis* (FO) ($m\approx 0.02$ mg), and small wasps *E. formosa* (EF) ($m\approx 0.02$ mg). The mass of those miniature insects ranges from 0.02 mg to 0.25 mg. From the literature, they obtained the power requirement data for medium and large insects with mass ranging from about 1 mg (fruitfly *D. melanogsater*) to 1600 mg (Hawkmoth *Manduca Sexta*). Now, with the power requirements of the miniature insects and those of the medium and large insects available, they can examine how the power requirement changes with size across the "full size range" of insects. Figure 12, adapted from Lyu and Sun (2021), compares the mass-specific powers of the insects (the nine species of medium and large insects considered in earlier literature and the six species of miniature insects considered by Lyu and Sun). More recently, Farisenkov *et al.* (2022) computed the power requirement of an even smaller miniature insect, the tiny beetle *P. placentis* ($m\approx 0.0024$ mg). The results of this tiny beetle, $(P^*)_1 \approx (P^*)_2 = 28$ W·kg$^{-1}$, is also added to Fig. 12.

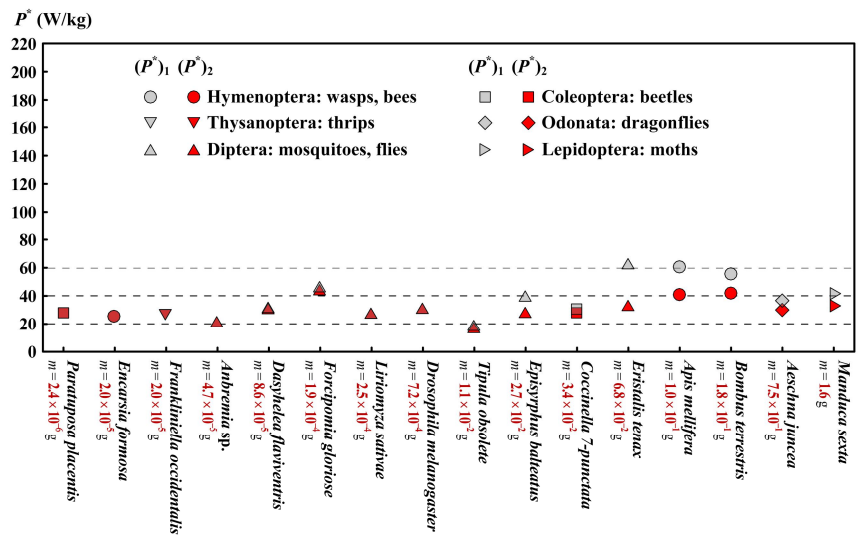

FIG. 12. Mass-specific powers in cases of 0% and 100% elastic energy storage for sixteen species of insects. Modified from Lyu and Sun (2021).

In Fig. 12, the mass of the smallest insect (tiny beetle *P. placentis*) is 0.0024 mg and that of the largest insect (hawkmoth *M. sexta*) is 1648 mg. There is a six orders of magnitude difference in mass. As seen from Fig. 12, even with the very large difference in size, these insects have relatively small difference in their mass specific power: $(P^*)_2$ of these insects approximately vary in the range of 20–40 W/kg, and $(P^*)_1$ in the range of 20–60 W/kg. Compared with the difference in their sizes, the difference in their mass-specific power is negligible, i.e., the power consumption of an insect in hovering flight is approximately proportional to its mass. Assuming that the mean power per unit muscle mass is the same under the same type of muscle, the above size/specific-power relation indicates that the ratio of flight-muscle mass to insect mass is approximately the same for different sized insects.

**B. Flight stability**

Dynamic flight stability (inherent or passive stability) is of great importance in the study of biomechanics of insect flight (Taylor and Thomas, 2003; Sun and Xiong, 2005; Hedrick, Cheng and Deng, 2009; Ristroph *et al*. 2013; Cheng and Deng, 2011). It is the basis for studying flight control, because the passive stability of a flying system represents the dynamic properties of the basic system, such as which degrees of freedom are unstable, how fast the instability develops, which variables are observable, and so on.

In the study of flight stability in insects, the averaged model is commonly used (e.g. Taylor and Thomas, 2003): the insect is treated as a rigid body of 6 degrees of freedom and the action of the flapping wings is represented by the wingbeat-cycle-average forces and moments. Thus the equations of motion of the insect are the same as those of an airplane. The equations of motion are linearized by approximating the body's motion as a series of small disturbance from a steady, symmetric reference flight condition. As a result of the linearization, the longitudinal and lateral small disturbance equations are de-coupled and can be solved separately. The longitudinal (or lateral) small disturbance equations are a system of four linear differential equations. Let us denote the system matrix of the longitudinal (or lateral) equations as *A* (the values of the elements of *A* can be computed by numerically solving the Navier-Stokes equations or by experimentally measuring the aerodynamic forces and moments of the flapping wings. Then, the central elements of the solutions for the flight stability problem are the eigenvalues of *A*. Being a forth order matrix, *A* has four eigenvalues ($\lambda_1$, $\lambda_2$, $\lambda_3$, $\lambda_4$). A real eigenvalue, or a conjugate pair of complex eigenvalues, represent a natural mode of the system. The motion of the flying body after an initial deviation from its reference flight is a linear combination of the natural modes. In a natural mode, the real part of the eigenvalue determines the time rate of growth of the disturbance quantities. A positive (negative) real eigenvalue will result in exponential growth (decade) of each of the disturbance quantities, so the corresponding natural mode is dynamically unstable [termed unstable divergent (stable subsidence) mode]. For an unstable divergent mode, the time to double the starting value ($t_d$) is given by

$$t_d = 0.693/\lambda \tag{7}$$

A pair of complex conjugate eigenvalues, e.g., $\lambda_{1,2}= s\pm\omega i$, will result in oscillatory time variation of the disturbance quantities with $\omega$ as its angular frequency. The motion grows when $s$ is positive (termed unstable oscillatory mode). The time to double the oscillatory amplitude is

$$t_d = 0.693/s. \tag{8}$$

Until recently, studies on flight stability were for medium and large insects (e.g. Taylor and Thomas, 2003; Cheng and Deng, 2011; Wu and Sun, 2012); the mass of these insects ranges from about 1600 mg (hawkmoths) to about 1 mg (mosquitoes and fruitflies). This covers the mass range of many winged insects, except that of the miniature insects whose mass is more than one order of magnitude smaller. One reason for the absence of the stability analysis of miniature insects was that their wing kinematical and morphological data were not available. Recently, the wing data for some miniature insects were measured (Lyu, Zhu and Sun, 2019; Farisenkov *et al.*, 2022) and the stability properties of these miniature insects can be calculated. As a first step, Lyu and Sun (2022) considered the longitudinal flight stability problem of two hovering miniature insects: vegetable leaf miner *L. sativae* ($m\approx0.25$ mg) and gall midge *Anbremia sp.* ($m\approx0.05$ mg). They found that for each of the two miniature insects, there is a pair of complex eigenvalues, $\lambda_1$ and $\lambda_2$, which have a positive real part, and there are two negative real eigenvalues ($\lambda_3$ and $\lambda_4$), one with a large magnitude and the other with a small magnitude. Therefore, the longitudinal motion has three natural modes: an unstable oscillatory mode, a stable fast subsidence mode and a stable slow subsidence mode. Owing to the unstable mode, the longitudinal motions of the gall midge and the vegetable leaf miner are unstable.

By comparing with the results of the medium and large insects, they pointed out that the modal structure of the two miniature insects is the same as that of the medium and larger insects: having an unstable oscillatory mode, a stable fast and a stable slow subsidence modes. That is, although the insects considered have a 30000-fold difference in mass (mass of the gall midge, 0.05 mg; that of the hawkmoth, 16000 mg), they have same modal structure. Because of the unstable mode, the hovering flight of insects of all sizes considered is passively unstable. This means that the flight must be actively controlled to be stable: the insects need to constantly react to their surroundings and adjust their wing motion in order to keep from tumbling. The response time of the nervous system needs to be fast enough to react and keep the unstable mode from growing too large. Therefore, the growth rate of the unstable mode is of great interest. The time to double the initial values of disturbances, $t_d$ [Eq. (7) or (8)] represents the growth rate of instability. Lyu and Sun (2022) calculated the values of $t_d$ of the two miniature insects, and they also calculated the values of $t_d$ for the other medium and larger insects using the values of $\lambda_1$ and $\lambda_2$ from previous studies. The value of $t_d$ for the insects (twelve species) are plotted in Fig. 13. An approximate analytical expression of $t_d$ as a function of $m$ was derived by Lyu and Sun (2022); it shows that $t_d$ is proportional to the 0.17 power of $m$ (also plotted in Fig. 13). That is, as $m$ becomes smaller, $t_d$ decreases (i.e., the instability becomes faster). This means that miniature insects need a faster nervous system to control the instability than larger insects. For example, the response time, represented by $t_d$, of the miniature insect, gall midge, whose $m \approx 0.05$ mg, needs to be faster by approximately 7 times than that of the large insect hawkmoth ($m \approx 1600$ mg).

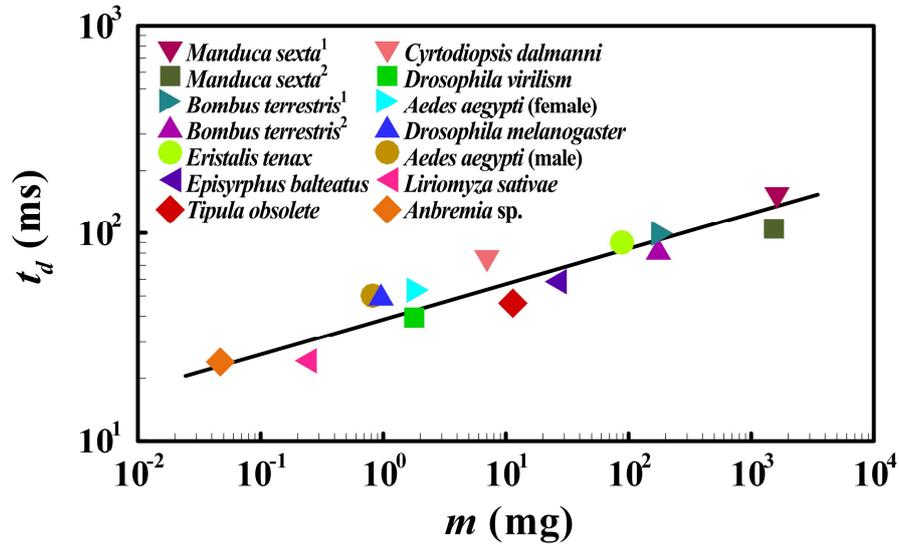

FIG. 13. The relationship between $t_d$ and $m$ of twelve species of insects. Adapted from Lyu and Sun (2022).

## VII. CONCLUDING REMARKS

In order to overcome the very large viscosity effects, miniature insects must employ different wing motion and aerodynamic mechanisms from those of the medium and large insects. Current data show the following changes. The planar upstroke commonly used by medium and large insects changes to a U-shaped upstroke. As insect size (represented by its mass $m$) becomes smaller, deeper and deeper U-shaped upstroke is employed. When $m$ decreases to about 0.02 mg, the downstroke also becomes U-shaped, but the U-shape is shallower. When $m$ decreases to about 0.0024 mg, the downstroke also becomes deep U-shaped. The features of a deep U-shape upstroke or downstroke are: in its first phase, the wings rapidly accelerate downward at nearly 90° angle of attack, "smashing" the air and producing a large drag or vertical force ('impulsive rowing' mechanism), whereas in its second phase, the wings are close to each other and move slowly upward at zero angle of attack, "slicing" through the air and producing a very small negative vertical force; thus the deep U-shape up- or downstroke gives a large mean vertical force. It should be noted that the above results were obtained from only eight species of insects whose wing kinematics have so far been measured (fruitfly *D. virilis*, m≈0.72mg; vegetable leafminer *L. sativae*, 0.25 mg; biting midge *F. gloriose*, 0.2 mg; biting midge *D. flaviventris*, 0.08 mg; gall midge *Anbremia* sp., 0.05 mg; thrip *F. occidentalis*, 0.02 mg; tiny wasp *E. formosa*, 0.02 mg and tiny beetle *P. placentis*, 0.0024mg). It is of great importance to measure the wing motion of more miniature insects and see if the above trend of variation in flapping pattern is general, and if there are other novel flapping patterns and new aerodynamic mechanisms.

The wings of many miniature insects are bristled rather than membranous. So far, the numerical studies on bristled wings are based on the Navier-Stokes equations with no-slip condition on the bristle surface; the experimental studies use dynamically scaled models moving in mineral oil. That is, in both the numerical and experimental studies, the flow is assumed to be in the continuum regime. It is known that for the flow to be in the continuum regime, the Knudsen number ($Kn$) needs to be small [$Kn$ is the ratio of the molecular mean free path of the air ($\lambda$) to a characteristic length of the geometry ($L_c$), e.g., the diameter of a cylinder]. Conventionally, when $Kn \lesssim 0.001$, the flow is well in the continuum regime and the Navier–Stokes equations are valid and the no-slip boundary condition can be used. When $Kn \gtrsim 0.001$, there will be rarefied-gas effects on the flow, but if $0.001 \lesssim Kn \lesssim 0.1$, the rarefied-gas effects mainly appear in the areas very near the body boundary and the major phenomenon is that the no-slip condition is violated: the flow slips on the boundary (the flow regime of $0.001 \lesssim Kn \lesssim 0.1$ is referred to as slip regime) (Dyson *et al.*, 2008). As already noted by some researchers (Liu and Aona, 2009; Santhanakrishnan, *et al.*, 2014), because of the small geometrical length scales of the bristled wings, *Kn* may be larger than 0.001 and there may be rarefaction effects. At 20°C and standard pressure, using the formula given in literature (see, e.g., Gu *et al.*, 2019), *λ* of the air is calculated as 0.065 μm. The diameter of the bristles (*D*) of the bristled wings, for which measurements have been made, ranges from about 0.65 μm to 2.4 μm (Jones *et al.*, 2016; Farisenkov *et al.*, 2022; Jiang *et al.*, 2022 ). Using *D* as the characteristic length, *Kn* ranges from 0.027 to 0.1; the flow is well within the slip regime. The rarefaction effects need to be studied in future research.

In the study of the tiny bristle-winged beetle *P. placentis*, the mass of the equivalent membranous wing was estimated by using the wing thickness of some of larger membranous-winged insects. The estimated mass could be too large; furthermore, in the estimating the moment of inertia of the equivalent membranous wing, a uniform wing thickness was used, this may over estimate the moment of inertia. Over-estimated mass and moment of inertia of the equivalent membranous wing may give over-estimated inertia power. Therefore, weather or not a bristle-winged tiny insect needs much less power than a membranous-winged insect of the same size, is still not clear. Study using same sized miniature bristled and membranous winged insects may clear this point.

Dynamic flight stability analysis has been made for two species of miniature insects in hovering flight, the vegetable leaf miner *L. sativae* ($m \approx 0.25$ mg) and the gall midge *Anbremia sp.* ($m \approx 0.05$ mg). The analysis shows that the longitudinal modal structure of the two miniature insects is the same as that of the medium and large insects: there is an unstable oscillatory mode, a stable fast and a stable slow subsidence modes. The flight is unstable because of the unstable mode. The difference between a smaller insect and a larger one is that the instability of the smaller insect grows faster than that of the larger one. The stability analysis has been based on the averaged-model theory and treats the flight as a fixed-point equilibrium. Because of the periodically varying aerodynamic and inertial forces of the flapping wings, a hovering or constant-speed flying insect is a cyclically forcing system, and generally the flight is not in a fixed-point equilibrium, but in a cyclic-motion equilibrium. The averaged-model theory gives good results for insects with relatively small body oscillation at wingbeat frequency; but for some insect with relatively large body oscillation at wingbeat frequency (e.g, large moths and butterflies), cyclic-motion stability analysis is required (Wu and Sun, 2012). The wing and body motions of eight species of miniature insects have been video recorded and measured (Lyu, Zhu and Sun, 2019; Farisenkov *et al.*, 2022). From the video recordings, it can be observed that body pitch oscillation is quite large for the smallest insect of the eight species, the tiny beetle *P. placentis*: its body pitch angle variation in the oscillation is approximately 30°. For a miniature insect like this, one may need to treat the flight as a cyclic-motion equilibrium, and use the Floquet theory or use numerical simulation by solving the complete equations of motion coupled with the Navier-Stokes equations, to analyze the flight stability. This is a very interesting and important future work, because knowing the passive stability properties of an insect is the basis for studying its stabilization of flight.

Flight control is an important part of insect flight; without control insects cannot really fly. Flight control involves coupling of the "inner" control systems (sensory system and neuron-motor control system) and the "outer" dynamics (passive stability, wing-motion change and aerodynamic-force change). There are broadly two types of flight control. One is stabilization control, which is used to stabilizing the flight (keep the disturbances from growing during hovering or constant-speed flight). The other is maneuver control, which is used to generate aerobatics, such as fast changing flight direction, recovering from upside down falling and landing on a ceiling. For medium and large insects, interesting and important studies have been made on both stabilization control (e.g. Ristroph *et al.*, 2010; Cheng, Deng and Hedrick, 2011; Ristroph *et al.*, 2013; Windsor *et al.*, 2014) and maneuver control (e.g. Haselsteiner, Gilbert and Wang, 2014; Liu *et al.*, 2019; Wang, Melfi and Leonardo, 2022). For miniature insects, however, study on flight control is so far absent. Research in this direction is strongly recommended.

**Acknowledgments**

The author thanks X. Cheng, Y. Z. Lyu, Z. H. Zhu, L. G. Liu, W. Y. Cai and W.

J. Liu for their help, and Y. l. Zhang, Y. P. Liu and D. Du for useful comments. The author acknowledges financial support from the National Science Foundation of China through Grants No. 11832004 and No. 11721202.**References**

Aono, H., F. Liang and H. Liu, 2008, "Near- and far- field aerodynamics in insect hovering flight: and integrated computational study," J. Exp. Biol., **211**, 239.

Arora, N., A. Gupta, S. Sanghi, H. Aono ans W. Shyy, 2014, "Lift–drag and flow structures associated with the 'clap and fling' motion," Phys. Fluids, **26**, 071906.

Barta, E. and D. Weihs, 2006, "Creeping flow around a finite row of slender bodies in close proximity," J. Fluid Mech. **551**, 1.

Batchelor, G. K., 1970, *An introduction to fluid dynamics*, Cambridge University Press.

Bomphrey, R. J., T. Nakata, N. Philips, and S. M. Walker, 2017, "Smart wing rotation and trailing-edge vortices enable high frequency mosquito flight." Nature **544**, 92–95.

Bomphrey, R. J., R. B Srygley, G. K. Taylor and A. L. R. Thomas, 2002, "Visualizing the flow around insect wings," Phys. Fluids **14**, S4.

Cheng, B. and X. Y. Deng, 2011, "Translational and rotational damping of flapping flight and its dynamics and stability at hovering", IEEE Trans. Robot. **27**, 849.

Cheng, B., X. Deng, and T. L. Hedrick, 2011, "The mechanics and control of pitching manoeuvres in a freely flying hawkmoth (Manduca sexta)," J. Exp. Biol. 214, 4092.

Cheng, X. and M. Sun, 2016, "Wing-kinematics measurement and aerodynamics in a small insect in hovering flight," Sci. Rep. **6**, 25706.

Cheng, X. and M. Sun, 2018, "Very small insects use novel wing flapping and drag principle to generate the weight-supporting vertical force," J. Fluid Mech. **855**, 646.

Cheng, X. and Sun, M., 2019, "Revisiting the clap-and-fling mechanism in small wasp *Encarsia formosa* using quantitative measurements of the wing motion," Phys. Fluids **31**, 101903.

Cheng, X. and M. Sun, 2021, "Wing kinematics and aerodynamic forces in miniature insect *Encarsia formosa* in forward flight," Phys. Fluids 33, 021905.

Dickinson, M. H., F. O. Lehman, S. P. Sane, 1999, "Wing rotation and the31